\documentclass{elsart}
\usepackage{graphicx,natbib,amssymb,lineno,longtable}
\journal{New Astronomy}
\def\astrobj#1{#1}
\begin{document}
\begin{frontmatter}

\title{First robotic monitoring of a lensed quasar: intrinsic variability 
of \astrobj{SBS 0909+532}\thanksref{LivTri}}
\thanks[LivTri]{Based on observations made with the Liverpool Telescope 
operated on the island of La Palma by Liverpool John Moores University 
in the Spanish Observatorio del Roque de los Muchachos of the 
Instituto de Astrofisica de Canarias with financial support from the 
UK Science and Technology Facilities Council. This research is part of 
the GLENDAMA project (http://grupos.unican.es/glendama/) and a wider
Euroasiatic programme on optical monitoring of gravitationally lensed
quasars.}
\author[UC]{L. J. Goicoechea\corauthref{cor}},
\corauth[cor]{Corresponding author.}
\ead{goicol@unican.es}
\author[IRE]{V. N. Shalyapin},
\ead{vshal@ire.kharkov.ua}
\author[SAI]{E. Koptelova},
\ead{koptelova@xray.sai.msu.ru}
\author[US]{R. Gil-Merino},
\ead{rodrigo@physics.usyd.edu.au}
\author[IAKhNU]{A. P. Zheleznyak},
\ead{zheleznyak@astron.kharkov.ua}
\author[UC]{A. Ull\'an}
\ead{aurora.ullan@postgrado.unican.es}

\address[UC]{Departamento de F\'{\i}sica Moderna, Universidad de Cantabria, 
Avda. de Los Castros s/n, 39005 Santander, Spain}
\address[IRE]{Institute for Radiophysics and Electronics, National Academy 
of Sciences of Ukraine, 12 Proskura St., Kharkov 61085, Ukraine}
\address[SAI]{Sternberg Astronomical Institute, Universitetski pr. 13, 119992 
Moscow, Russia}
\address[US]{Institute of Astronomy, School of Physics, The University of 
Sydney, NSW 2006, Australia}
\address[IAKhNU]{Institute of Astronomy of Kharkov National University, 
Sumskaya 35, 61022 Kharkov, Ukraine}

\begin{abstract}
To go into the details about the variability of the double quasar \astrobj{SBS 0909+532}, we designed a monitoring 
programme with the 2 m Liverpool Robotic Telescope in the $r$ Sloan filter, spanning 1.5 years from 2005 
January to 2006 June. The $r$--band light curves of the A and B components, several cross--correlation 
techniques and a large number of simulations (synthetic light curves) lead to a robust delay $\Delta 
t_{BA}$ = $-$ 49 $\pm$ 6 days (1$\sigma$ interval) that agrees with our previous results (the B component 
is leading). Once the time delay and the magnitude offset are known, the magnitude-- and time--shifted 
light curve of image A is subtracted from the light curve of image B. This difference light curve of \astrobj{SBS 0909+532} 
is consistent with zero, so any possible extrinsic signal must be very weak, i.e., the observed 
variability in A and B is basically due to observational noise and intrinsic signal. We then make the 
combined light curve and analyse its statistical properties (structure functions). The structure function 
of the intrinsic luminosity is fitted to predictions of simple models of two physical scenarios: accretion 
disc instabilities and nuclear starbursts. Although no simple model is able to accurately reproduce the 
observed trend, symmetric triangular flares in an accretion disc seems to be the best option to account 
for it.
\end{abstract}


\end{frontmatter}

\section{Introduction}

\astrobj{SBS 0909+532} consists of two components (two--image gravitationally lensed quasar) separated by about 
1.1$''$ \citep{Koc97,Leh00,Lub00,Osc97a}. In optical frames taken at normal seeing conditions with 
relatively short exposure times, the lensing elliptical galaxy is undetectable \citep[e.g.][]{Ull06}. 
Thus, a simple photometric model (with only two close point--like sources) is able to describe the 
whole crowded region associated with the quasar components. The first resolved light curves of \astrobj{SBS 0909+532} 
were presented by \citet{Ull06}, who derived accurate fluxes in the $R$ Johnson--Cousins--Bessel 
filter from a multisite observing campaign in 2003. These first $R$-band records were used to obtain the 
time delay between both components. From two different techniques and 1000 repetitions of the experiment 
(synthetic light curves based on the observed records), \citet{Ull06} reported a delay $\Delta t_{BA}$ 
ranging from $-$ 41 to $-$ 56 days ($\geq$ 90--95\% confidence), where the minus sign means that the 
intrinsic signal is observed first in the faintest component (B) and later in the brightest component 
(A). However, the first variability study had some weak points. The $R$--band light curves did not permit 
to rule out positive delays fairly, and a negative interval [$-$ 90, 0] days was considered in the 
estimation of uncertainties (component B leading component A). Moreover, there was a relatively poor 
overlap between the A and B records, when the A light curve was shifted by the best solutions of the 
time delay and the magnitude offset. 

The light curves of the two components of a double quasar at a given wavelength, provide extremely 
valuable astrophysical information. As the ray paths are different for different components, the 
corresponding traveltimes will not agree with each other: it appears a time delay between the observed 
components, which is directly related to the present expansion rate of the Universe (Hubble constant) 
and the mean surface density of the lensing galaxy \citep[e.g.][]{Koc04,Ref64}. To tackle the determination 
of $H_0$ and $<\kappa>$, one needs to measure the time delay and the basic parameters of the gravitational 
mirage, i.e., redshifts and image positions, where the relevant image positions are the positions with 
respect to the centre of the lensing galaxy (the matter/energy content of the Universe plays also a role). 
Apart from the time delay, a magnitude offset can be also derived from the comparison of both brightness 
records. Thus, the possible extrinsic variability (due to intervening objects, e.g., microlenses or dusty 
clouds in the lensing galaxy) may be detected through the difference light curve ($DLC$): the difference 
between the light curves of the components, when the time delay and the magnitude offset are taken into 
account properly \citep{Sch98}. While the $DLC$s of \astrobj{QSO 0957+561} do not show evidence for extrinsic 
variability \citep{Gil01,Sch98,Wam98}, the $DLC$s of other lens systems seem to indicate the existence of 
extrinsic gradients and fluctuations \citep[e.g.][]{Par06}. 

When the difference signal is consistent with zero, the variability observed in both components 
can be attributed to observational noise and intrinsic phenomena, i.e., physical processes in the 
source quasar. In this hypothetical case there is a unique opportunity to study the intrinsic signal of 
a distant quasar \citep[e.g.][]{Kaw98}. The light curves of the two components can be combined 
to produce one better--sampled record, and this combined light curve ($CLC$) can be used to carry out 
statistical analyses. \citet{Kaw98} analysed the logarithmic slope of the structure function 
\citep[e.g.][]{Sim85} of a $CLC$ of \astrobj{QSO 0957+561}, which (slope) is in reasonable agreement with 
the predictions by an accretion disc--instability model. They remarked one important caveat about the 
methodology to discriminate between different scenarios, which is based on the logarithmic slope of the 
structure function of the intrinsic luminosity for nuclear starbursts and accretion disc instabilities. 
The analysis only incorporated two simple models, i.e., a standard starburst model \citep{Are97} leading 
to relatively high slopes and a cellular--automaton disc--instability model \citep{Min94} producing the 
smallest slopes, so the conclusions could be biased if the simple models do not describe the actual 
behaviour of the two physical scenarios or the intrinsic signal is caused by several independent 
mechanisms. However, the comparison with simple analytical or numerical models represents an important 
first step to reveal the origin of the intrinsic signal. 

In order to get high--quality information about the variability of \astrobj{SBS 0909+532} at a red wavelength, we 
designed a monitoring programme with the world's largest fully robotic telescope. The 2 m Liverpool 
Robotic Telescope \citep{Ste04} at the Roque de los Muchachos Observatory (Canary Islands, Spain) was used 
from 2005 January to 2006 June to obtain nightly frames in the $r$ Sloan filter. In this paper we present 
the robotic programme and the corresponding light curves (Sect. 2). Section 3 is devoted to the time delay 
estimation from the records in the $r$ band. Unfortunately, the lens galaxy of \astrobj{SBS 0909+532} has a large 
effective radius, with a correspondingly low surface brightness \citep{Leh00}. These photometric properties 
complicate the goal of determining an accurate galaxy astrometry, and thus, accurate values of $H_0$ and 
$<\kappa>$. Even using the more recent optical frames taken with the Hubble Space Telescope (HST), the 
position of the centre of the galaxy has a large uncertainty. Consequently, we do not discuss the $H_0$ and 
$<\kappa>$ values, and focus on a careful analysis of the nature of the observed variability. In Sect. 4, 
we make the $DLC$ and look for possible extrinsic signal. The $CLC$ and its statistical properties 
(structure function) are analysed and interpreted in Sect. 5. Finally, in Sect. 6 we summarise our 
conclusions and put the results in perspective. 

\section{Observations and light curves}

The monitoring programme with the 2 m Liverpool Robotic Telescope (LRT) began in 2005 January. This
robotic project was carried out with the RATCam optical CCD camera. The field of view  and the pixel 
scale (binning 2$\times$2) were $\sim 4.6' \times 4.6'$ and 0.278 arcsec, respectively. 
Although we concentrate here on the observations in the red arm of the optical spectrum, i.e., the 
frames in the $r$ Sloan passband, the gravitationally lensed quasar was also monitored in the blue arm 
(via LRT in the $g$ Sloan passband). A chromatic ($gVrR$) multisite study will be presented further on. 
Our red subprogramme was optimized to get frames all nights when \astrobj{SBS 0909+532} is visible and there are 
no technical/atmospheric problems. Each useful night we usually obtained one $r$--band frame of the lens 
system (exposure time of 120 s). However, many nights in the 2005 October--December period we obtained 
two 120 s frames, so we are able to check formal photometric errors by comparing them with the 
artificial (untrue) intranight variabilities. In fact, the intranight scatters are the best non--biased 
estimators of the typical photometric errors. In Table \ref{t1} we include details about the whole $r$--band 
monitoring from 2005 January to 2006 June (month, frames and observing nights, sampling rate, and total 
exposure time). Note that this robotic project is quite cheap in observation time, since the total 
science time is only $\sim$ 18 ks = 5 hours. However, this kind of 1.5--year monitoring (see Table \ref{t1}) is 
organizationally very complex or impossible with a conventionally scheduled and operated telescope 
\citep{Ste00}.

\begin{table*}
\centering
\begin{minipage}{140mm}
\caption{Liverpool Robotic Telescope observations of \astrobj{SBS 0909+532} in the $r$ Sloan filter}
\label{t1}
\begin{tabular}{c c c c}
\hline
Month & Frames$\times$nights & Sampling rate (nights/week) & Total exposure time (s) \\
\hline
2005 January & 1$\times$2 & 0.5 & 240 \\
2005 February & 0 & 0 & 0 \\
2005 March & 1$\times$12 & 3.0 & 1440 \\
2005 April & 1$\times$2 & 0.5 & 240 \\
2005 May & 1$\times$7 & 1.7 & 840 \\
2005 June & 1$\times$1 & 0.2 & 120 \\
2005 July & occultation & --- & --- \\
2005 August & occultation & --- & --- \\
2005 September & occultation & --- & --- \\
2005 October & 1$\times$1 + 2$\times$6 & 1.7 & 1560 \\
2005 November & 1$\times$5 + 2$\times$11 & 4.0 & 3240 \\
2005 December & 1$\times$4 + 2$\times$12 & 4.0 & 3360 \\
2006 January & 1$\times$1 + 4$\times$1 & 0.5 & 600 \\
2006 February & 1$\times$8 + 2$\times$1 & 2.2 & 1200 \\
2006 March & 1$\times$16 + 2$\times$1 & 4.2 & 2160 \\
2006 April & 1$\times$11 + 2$\times$1 & 3.0 & 1560 \\
2006 May & 1$\times$13 & 3.2 & 1560 \\
2006 June & 1$\times$2 & 0.5 & 240 \\
\hline
\end{tabular}
\end{minipage}
\end{table*}

Basic instrumental reductions are applied to all RATCam frames before the data are passed to users. This 
incorporates bias subtraction, trimming of the overscan regions and flat fielding. After the basic 
pre--processing, cosmic ray rejection was applied to the frames. There are also bad pixel masks, which are 
kindly made available by the Angstrom project \citep{Ker06}, which is another gravitational lensing 
programme underway on the LRT. Additional details on the CCD, the Sloan filters and the data pipeline (LRT 
automatic pre--processing) can be found at http://telescope.livjm.ac.uk/.

\begin{figure}
\begin{center}
\includegraphics*[width=7cm,angle=0]{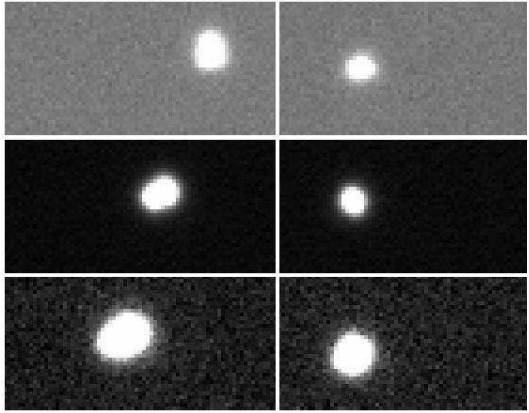}
\end{center}
\caption{Liverpool Robotic Telescope subframes of \astrobj{SBS 0909+532} (left panels) and "c" field star (right 
panels) in the $r$ Sloan filter. The top, middle and bottom panels correspond to exposures in 2005 March 
23 ($FWHM = 1.49''$), 2005 December 4 ($FWHM = 1.07''$) and 2006 March 18 ($FWHM = 1.15''$), 
respectively. All the subframes (panels) have similar size, and the brightness of the "c" star is similar 
to the brightness of the A component. The double quasar (\astrobj{SBS 0909+532}) roughly looks like two tangent 
stars.}
\label{f1}
\end{figure}

\begin{figure}
\begin{center}
\includegraphics*[width=7cm,angle=-90]{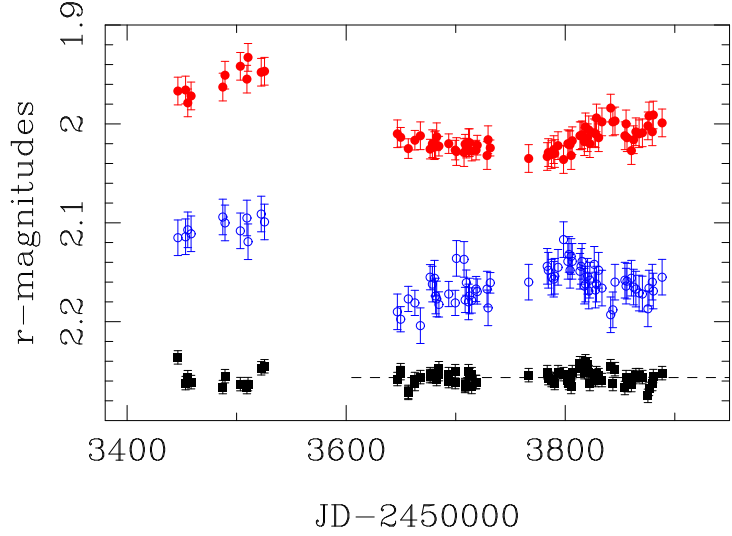}
\end{center}
\caption{Liverpool Robotic Telescope light curves of \astrobj{SBS 0909+532} in the $r$ Sloan filter. The filled 
circles are the fluxes $y_A$, shifted by + 0.50 mag, and the open circles are the fluxes $y_B$ (see main 
text). We also incorporate relative fluxes of the "a" star for comparison purposes. The filled squares 
represent the $y_a$ data, shifted by + 3.13 mag, and the discontinuous line is the fit to the fluxes 
corresponding to the 2005/2006 season (after day 3600). While the stellar fluxes are narrowly distributed 
around a constant flux, the A fluxes trace gradients and have a larger error, and the B fluxes have the 
largest uncertainty and scatter.}
\label{f2}
\end{figure}

A simple photometric model works well for \astrobj{SBS 0909+532} \citep[see Introduction and][]{Ull06}. Due to 
the faintness of the lensing galaxy, the whole crowded region can be described through only two close 
point--like sources (two components of the lensed quasar). Thus, we apply a PSF fitting method to a 
representative subset of LRT optical frames verifying some elemental conditions (e.g., the telescope 
pointing was accurate enough so that the lens system is included in the field of view, there is no a strongly
degraded signal, etc). The subset contains 92 frames, so 60\% of the robotic exposures in Table \ref{t1} are 
used to make the light curves. In Fig. \ref{f1} we show subframes corresponding to three successful exposures: 2005 
March 23 (top panels), 2005 December 4 (middle panels) and 2006 March 18 (bottom panels). The double quasar
(left panels) appears as an extended structure similar to two tangent stars (see the right panels including 
the "c" field star). Following the notation of \citet{Ull06}, we obtain measurements of $y_A = m_A - m_b$, 
$y_B = m_B - m_b$, $y_a = m_a - m_b$ and $y_c = m_c - m_b$, i.e., we compute relative magnitudes using the 
"b" star as the reference object \citep[the lens system is inside the triangle defined by the "a--c" 
stars, e.g.][]{Koc97}. Both "a" and "b" are non--variable field stars, and we expect a constant behaviour 
of $y_a$. To check the trend of $y_a$, we focus on the 69 relative fluxes corresponding to the 2005/2006 
season: continuous monitoring from 2005 October to 2006 June. Using the formal photometric errors, the trend 
is not consistent with a constant because the reduced chi--square value is large ($\chi^2$ = 6.27). The 
formal uncertainties do not seem to include all the sources of error, and we need a non--biased estimator 
of uncertainties. This is not a problem in our project, since we have several nights with two exposures at 
different times. Taking into account the deviations in $y_a$ for nights with two data (intranight 
deviations: $\delta_k = y_a(t_k + \Delta t) - y_a(t_k)$, $\Delta t \leq$ 8 hours, $k = 1,...,N$), it is 
easy to obtain a standard intranight deviation $\sigma_a = \sqrt{\sum_{k=1}^{N} \delta_k^2 /(N-1)}$ = 6.8 
mmag. Assuming this standard deviation as a typical error (i.e., the intranight variability is not real, 
but the tool to estimate the true photometric uncertainty) and re--doing the fit to a constant, we derive 
a very reasonable $\chi^2$ value of 1.12 (see the filled squares and the discontinuous line after day 3600 
in Fig. \ref{f2}). Therefore, the $m_a - m_b$ record in the 2005/2006 season is clearly consistent with a constant 
behaviour and there is a fair way to infer non--biased uncertainties of the quasar fluxes $y_A$ and $y_B$ 
(typical errors from standard intranight deviations).  

\begin{figure}
\begin{center}
\includegraphics*[width=7cm,angle=-90]{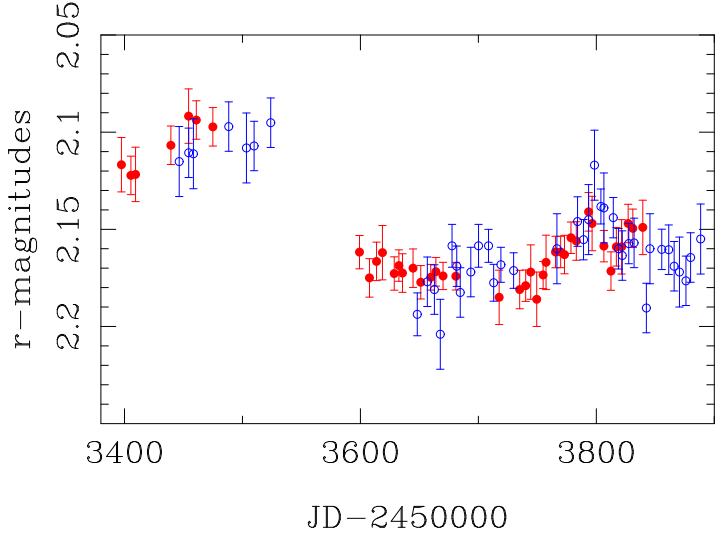}
\end{center}
\caption{Grouped light curves of \astrobj{SBS 0909+532}. We grouped the individual quasar fluxes (see Fig. \ref{f2}) 
within 3--day intervals. The filled circles are the fluxes $y_A$, shifted by $-$ 49 days and + 0.65 mag
(see section 4), and the open circles are the fluxes $y_B$.}
\label{f3}
\end{figure}

From intranight variabilities in the 2005/2006 season, we derive self--consistent uncertainties: 
$\sigma_a \sim$ 7 mmag $< \sigma_c \sim$ 10.5 mmag $< \sigma_A \sim$ 14 mmag $< \sigma_B 
\sim$ 18 mmag. We remark that the "c" star and the A component have a similar brightness, but A is 
placed in a crowed region and reasonably $\sigma_A$ is larger than $\sigma_c$. While the stellar relative 
fluxes are inferred with true uncertainties less than or equal to 10 mmag, the quasar relative fluxes 
have true errors in the 10--20 mmag range. Once non--biased errors in $y_A$ and $y_B$ are determined from 
intranight variations, we group the quasar fluxes each night including more than one exposure. In the 
2005/2006 season, the quasar light curves are characterized by a mean sampling rate of two points every 
week. The LRT was offline in 2005 September (for two weeks) for engineering work which included 
realuminisation of the primary mirror. Thus, our whole original light curves (incorporating data before 
2005 September) are affected by this maintenance work, and we must correct the realuminisation offsets, 
i.e., the instrumental offsets between 2004/2005 and 2005/2006 seasons. There is no problem with $y_a$,
since we can easily estimate the difference between the average fluxes before and after day 3600. However,
the quasar components are variable objects, and we need independent brightness records to correct the
offsets in both $y_A$ and $y_B$. A simultaneous monitoring in the $R$ band at Mt. Maidanak, Uzbekistan, 
is used to derive the realuminisation biases in the LRT records. Maidanak (1.5 m AZT--22 Telescope) quasar 
fluxes are compared with LRT brightnesses at separations less than or equal to 3 days, which leads to 
instrumental offsets of about 70 and 10 mmag in $y_A$ and $y_B$, respectively. As the $R - I$ and $B - R$
colours of the "b" star are close to the colours of the B component (similar spectra), it is expected a 
very small correction in the differential curve $y_B$. For this record $y_B$, we just obtain an offset 
(slightly less than 10 mmag) that agrees with theoretical predictions. The whole final brightness records 
in the $r$ band are depicted in Fig. \ref{f2}. In this figure, we plot the A light curve (filled circles), shifted 
by + 0.50 mag, and the B light curve (open circles). To compare quasar fluxes against stellar brightnesses, 
relative fluxes of the "a" star (filled squares), shifted by + 3.13 mag, are also shown in Fig. \ref{f2}. The LRT 
brightnesses in the $r$-band (78 data of each component) have errors above 10 mmag, which could complicate 
some analyses. Thus, to improve the situation we also group fluxes within 3--day intervals. The new data 
set only contains 41 fluxes of each component, but both mean errors are reduced to the 10--mmag level: 
$<\sigma_A> \sim$ 10 mmag and $<\sigma_B> \sim$ 13 mmag. In Fig. \ref{f3} we show the new grouped light curves, 
i.e., the A curve (filled circles) and the B curve (open circles). Now, to easily compare both trends, the 
A record is properly shifted in time and magnitude ($-$ 49 days and + 0.65 mag; see sections 3--4). The 
individual calibrated quasar fluxes are presented in Table \ref{t2}. We use the SDSS flux of the "b" star in the
$r$ band ($m_b$ = 14.87 mag) to get apparent magnitudes of both components\footnote{The SDSS Web site is 
http://www.sdss.org/.}.  

\section{Confirmation of the time delay}

If we concentrate our attention in Fig. \ref{f3}, the A and B light curves show a decline larger than 50 
mmag as well as a 50--mmag event around day 3800 (A is shifted in time and magnitude). Thus, although 
there is an important gap in the LRT monitoring (due to occultation of the lens system in 2005 
July--September), these two features are promising tools to confirm our previous time delay estimation 
(see Introduction). To calculate the time delay between both components of \astrobj{SBS 0909+532}, we use 
different cross--correlation techniques. First, we focus on the grouped light curves and the discrete
cross--correlation function ($DCF$). The $DCF$ was introduced by \citet{Ede88} and has been extensively 
used in delay studies \citep[e.g.][]{Gil02,Ofe03,Osc97b}. Here, the $DCF$ of \astrobj{SBS 0909+532} is evaluated 
every 1 day in the region from $-$ 200 to + 100 days, so we analyse a wide range of lags including both 
positive and negative values. The $DCF$ is binned in 2$\alpha$ day intervals centered at the lags. To 
work with a reasonable time resolution, we only take into account $\alpha$ values less than or equal to 
10 days, i.e., bins with width $\leq$ 20 days. For $\alpha \leq$ 4 days, the $DCF$ is very noisy, whereas 
for $\alpha$ = 8--10 days, the main peak of the $DCF$ is significantly reduced with respect to the 
expected value of 1. The more interesting results are derived from $\alpha$ = 5--7 days, and we choose the 
intermediate bin ($\alpha$ = 6 days) as the most suitable one. For $\alpha$ = 6 days, there is a maximum 
at $-$ 50 days ($DCF_{max} \sim$ 0.9). Apart from the main peak around the maximum (delay--peak), there 
are other secondary peaks at negative and positive lags. These secondary structures have an amplitude of 
about 0.5, i.e., they are clearly smaller than the main feature. No correlation ($DCF \sim$ 0) or 
anticorrelation ($DCF <$ 0) is also found at the edges of the time lag--interval. For other $\alpha$ values, 
we also find maxima at or around $-$ 50 days.

As it was discussed by \citet{Leh92}, when the main fluctuations in the light curves have an intrinsic 
origin, the irregular delay--peak of the AB cross--correlation function should be closely 
traced by the symmetrical central peak (around a lag equal to zero) of the AA (or BB) autocorrelation 
function. Moreover, other features of the cross--correlation function around lags $T_1$, $T_2$,... will 
be closely reproduced in the autocorrelation function around lags $T_1 - \Delta t_{BA}$, $T_2 - \Delta 
t_{BA}$,..., respectively (with relation to A, we assume that B is delayed in $\Delta t_{BA}$). 
Therefore, if the shifted discrete autocorrelation function ($DAF$) is matched to the discrete 
cross--correlation function ($DCF$), one derives the time delay in a self--consistent way (note that 
the lag corresponding to the maximum of the $DCF$ is only a rough estimation of the delay). This 
self--consistent methodology is called the $\delta^2$ technique, and it has been successfully applied 
to some golden data sets (see next paragraph) for \astrobj{QSO 0957+561} \citep[e.g.][]{Goi98,Ser99}. 

Before applying the $\delta^2$ technique, we need to make a golden data set. A golden data set contains 
a variable and free from long gaps light curve A in a period [$t_i$, $t_f$], and a variable and free 
from long gaps light curve B in a period [$t_i + \Delta t$, $t_f + \Delta t$], where $\Delta t \sim 
\Delta t_{BA}$. As we know a rough initial estimation of the \astrobj{SBS 0909+532} delay (through the maximum of 
the $DCF$, see here above), we can use this value ($\Delta t$ = $-$ 50 days) to correct the unsuitable 
50--day edges. It is also clear that the records in Fig. \ref{f3} are variable. Therefore, the existence of
long gaps is the only difficulty to make a golden data set for \astrobj{SBS 0909+532}. Can we relax the condition
on the absence of gaps?. Simulations by \citet{Ser99} indicated that the $\delta^2$ method works even in 
the presence of relatively long gaps. Moreover, the main problem with gaps is the possible bias between 
the differences $y_A - <y_A>$ and $y_B - <y_B>$, which are the key pieces in the cross--correlation. In 
the periods of interest, the two light curves are sampled in different ways, so the direct averages of the 
fluxes $<y_A>_d$ and $<y_B>_d$ could lead to a bias between the differences $y_A - <y_A>_d$ and $y_B - 
<y_B>_d$. Thus we can obtain a quasi--golden data set provided that $<y_A>$ and $<y_B>$ are carefully 
determined. Instead of the direct averages of the grouped fluxes $y_A$ and $y_B$ in the periods of interest, 
the mean values $<y_A>$ and $<y_B>$ are inferred from a method that corrects for sampling bias. 

\begin{figure}
\begin{center}
\includegraphics*[width=7cm,angle=-90]{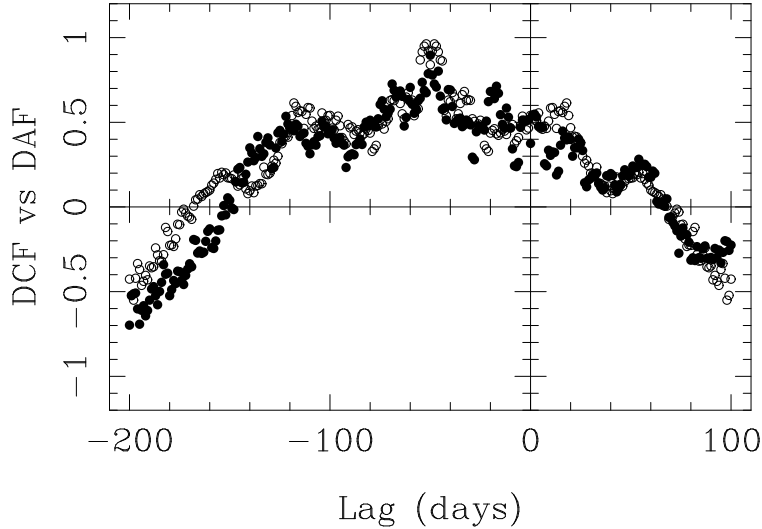}
\end{center}
\caption{Comparison between the $DCF$ (filled circles) and the $DAF$ shifted by $-$ 50 days (open 
circles). The $DAF$ is the average of the AA and BB autocorrelation functions. We use a quasi--golden
data set (see main text) and $\alpha$ = 6 days, where $\alpha$ is the semiwidth of the bins.}
\label{f4}
\end{figure}

\begin{figure}
\begin{center}
\includegraphics*[width=7cm,angle=-90]{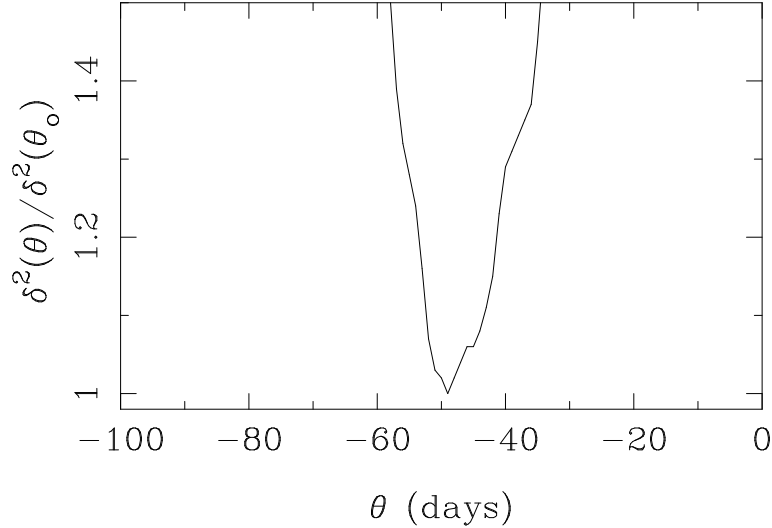}
\end{center}
\caption{Normalised $\delta^2$ function from LRT data in the $r$ band. The $\delta^2$ function is made
from the $DCF$ and $DAF$ in Fig. \ref{f4}.}
\label{f5}
\end{figure}

Once we have a quasi--golden data set, a comparison between the corresponding $DCF$ (filled circles) and 
$DAF$ (open circles) is plotted in Fig. \ref{f4}. The $DAF$ is the average of the AA and BB autocorrelation 
functions, and it is shifted by $-$ 50 days, i.e., the rough estimation of the delay. There is no 
important distortions in the features of the $DCF$ as compared with the features in the $DAF$, which 
strengths the realibility of a delay close to $-$ 50 days. Using $\alpha$ = 6 days, possible values of 
the time delay ($\theta$) versus the associated $\delta^2(\theta)$ values normalised by its minimum value 
$\delta^2(\theta_0)$, are also plotted in Fig. \ref{f5}. The $\delta^2(\theta)$ function is defined in Eq. (1) and 
Eq. (7) of \citet{Goi98} and \citet{Ser99}, respectively. In Fig. \ref{f5}, a relatively narrow peak centered 
on $-$ 49 days (best value of the delay) is derived. To obtain uncertainties, we follow an approach similar 
to that described by \citet{Ull06}. We make 1000 repetitions of the experiment, apply the $\delta^2$ 
minimization ($\alpha$ = 6 days) to each quasi--golden synthetic data set, and thus obtain 1000 best values 
of the delay. Through the distribution of delays, our 1$\sigma$ measurement is $\Delta t_{BA}$ = $-$ 49 
$\pm$ 7 days (69.8\% confidence interval). This $\delta^2$ result confirms the previous delay determination 
from Calar Alto and Maidanak frames in 2003. Moreover, the new measurement is very robust, since it is not a 
pre--conditioned estimation (a wide range from $-$ 200 to + 100 days is tested) and there is a significant 
overlap between the A and B records, when the A light curve is shifted by the best solution of the time delay 
(see Fig. \ref{f3}).

\begin{figure}
\begin{center}
\includegraphics*[width=7cm,angle=0]{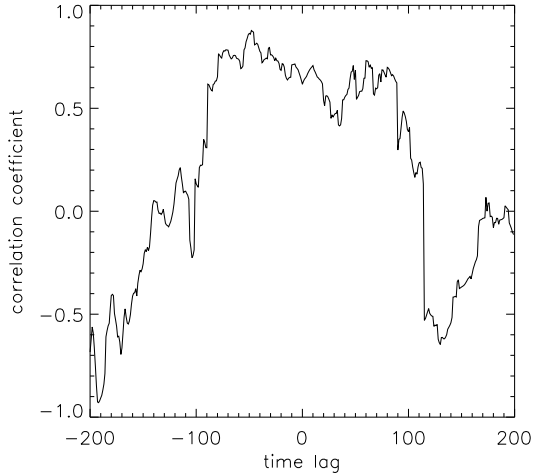}
\end{center}
\caption{Correlation coefficient associated with the $MCF$ method. We use the individual fluxes in 
Fig. \ref{f2} and 10--day bins in the component A.}
\label{f6}
\end{figure}

We also apply the modified cross--correlation function ($MCF$) technique \citep{Bes95,Okn97}. The $MCF$ 
combines properties of both standard cross--correlation functions: the interpolated cross-correlation 
function by \citet{Gas86} and the $DCF$ by \citet{Ede88}. This time we analyse the individual (non--grouped) 
fluxes in Fig. \ref{f2}, and after doing some tests, we use 10--day bins in the component A. When the $MCF$ is 
applied to our data in the lag interval [$-$ 200, + 200] days, the maximum correlation coefficient 
($R_{max}$ = 0.878) corresponds to a lag $T_{max}$ = $-$ 48 days. The correlation coefficient at different 
lags (days) is shown in Fig. \ref{f6}. We also check if the removal of some fluxes from the B record can influence 
the estimation of ($R_{max}$, $T_{max}$). The stability test is done in a blind way, i.e., we do not choose 
by eye the points to be dropped, but a systematic procedure is used to select those points. In a first step, 
the first 3 points are removed from the B light curve, then the second, third and fourth points are removed 
from, etc. From this 3--point systematic cleaning, almost all the $R_{max}$ values are close to 0.9 and 
$\sim$ 85\% of the iterations lead to $T_{max}$ = $-$ 49 days, in good agreement with the analysis of the 
whole data set. In order to estimate delay errors, we carry out simulations taking into account all the 
properties of the observed records: kind of variability, sampling and photometric errors \citep{Kop06}. 
About 1000 synthetic data sets leading to $R_{max} >$ 0.8 are considered here. Our 1$\sigma$ measurement 
is $\Delta t_{BA}$ = $-$ 49 $\pm$ 5 days (70.6\% confidence interval), which is practically identical to 
the 1$\sigma$ result from the $\delta^2$ method. We consider that the two techniques ($\delta^2$ and 
$MCF$) have similar quality. Thus, there is no a fair way to choose either 5 or 7 days as error, and we 
adopt $-$ 49 $\pm$ 6 days as the final 1$\sigma$ estimation, where $\pm$ 1 day is the uncertainty in the 
error. 

\section{Difference light curve}

From now on we take the LRT grouped fluxes in Fig. \ref{f3} as basic tools for discussing variability properties
in the $r$ filter. First, considering the time delay in section 3, we infer the time delay--shifted light
curve of A ($y_A^{TS}$). Here, $y_A^{TS}$ is the $y_A$ record shifted by $\Delta t_{BA}$ = $-$ 49 days in 
the horizontal direction. Then a mean offset $<y_B - y_A^{TS}>$ is computed in a direct way. This magnitude 
offset is close to + 0.65 mag. Second, the magnitude-- and time--shifted light curve of image A ($y_A^{MTS}$, 
where $y_A^{MTS}$ is the $y_A^{TS}$ curve shifted by the magnitude offset in the vertical direction) is 
subtracted from the light curve of image B ($y_B$). Both $y_A^{MTS}$ and $y_B$ are plotted in Fig. \ref{f3}. We 
thus obtain the difference light curve ($DLC$) $y_B - y_A^{MTS}$. To compute the mean offset as well as the 
$DLC$, the dates in the time shifted curves $y_A^{TS}$ and $y_A^{MTS}$ are taken as reference epochs. The 
$y_A^{TS}$ and $y_A^{MTS}$ fluxes are then compared to the averaged values of $y_B$ in bins with semiwidth 
$\alpha$ centred on the reference dates. 

\begin{figure}
\begin{center}
\includegraphics*[width=7cm,angle=-90]{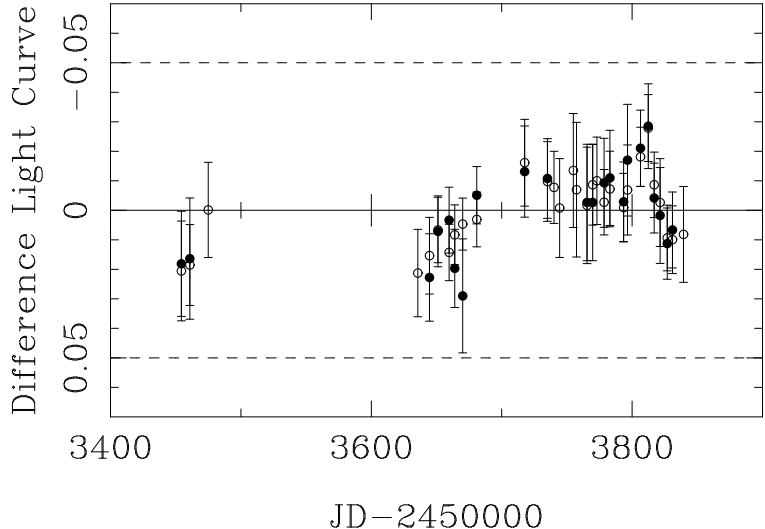}
\end{center}
\caption{Difference light curve of images B and A (shifted in time and magnitude) of the lensed quasar
\astrobj{SBS 0909+532}. To obtain the differences in $r$--magnitudes, we use bins in B with semiwidths $\alpha$ = 
6 days (filled circles) and $\alpha$ = 15 days (open circles). The $\pm$ 0.05 mag thresholds are depicted 
by discontinuous lines.}
\label{f7}
\end{figure}

The $DLC$ of \astrobj{SBS 0909+532} in the years 2005--2006 appears in Fig. \ref{f7}. We use two different $\alpha$ values:
6 days (filled circles) and 15 days (open circles), and both trends are consistent with each other. The 
differences (in $r$--magnitudes) do not exceed the $\pm$ 0.05 mag thresholds (discontinuous lines).
Moreover, there is no evidence in favor of the existence of events or gradients. The difference signal is
in apparent agreement with zero, i.e., Fig. \ref{f7} shows a noisy relationship $y_B$ = $y_A^{MTS}$. From a 
quantitative point of view, we can also estimate the goodness of representing the data with $y_B - 
y_A^{MTS}$ = 0 \citep[e.g.][]{Sch98}. The reduced chi--square values are 0.92 ($\alpha$ = 6 days) and 0.87 
($\alpha$ = 15 days), corroborating the good agreement between the light curves of the two images. Therefore, 
the observed variability in A and B is basically due to observational noise and intrinsic signal, although 
we cannot rule out the existence of a very weak extrinsic signal whose amplitude must be well below the 
noise level in the $DLC$. Our constraints on the possible microlensing (extrinsic) variability can be used 
to obtain information on the granularity of the matter in the lensing galaxy and the size of the source 
(Gil--Merino et al., in preparation). Although \citet{Med05} suggested the existence of an achromatic 
microlensing magnification of image B, a homogeneous microlensing pattern will not produce microlensing 
variations but a microlensing offset as part of the magnitude offset. Only an inhomogeneous microlensing 
pattern can be detected through the $DLC$.

\section{Combined light curve: origin of the intrinsic signal}

The $DLC$ presented in Sect. 4 is in clear agreement with the absence or an extremely low (undetectable) 
level of extrinsic signal. Therefore, in this section, we make the combined light curve ($CLC$) and 
interprete it as due to the observational noise and intrinsic phenomena. The combined photometry consists of 
both light curves $y_A^{MTS}$ and $y_B$, and this $CLC$ is plotted in Fig. \ref{f3} (filled and open circles). The 
data points in Fig. \ref{f3} show the global record $y_{qso}$ (in $r$--magnitudes) in a 500--day interval, with the 
data covering a period $P \sim$ 450 days (there is an unavoidable gap of about 50 days that is related to the 
annual occultation of the quasar). 

\begin{figure}
\begin{center}
\includegraphics*[width=7cm,angle=-90]{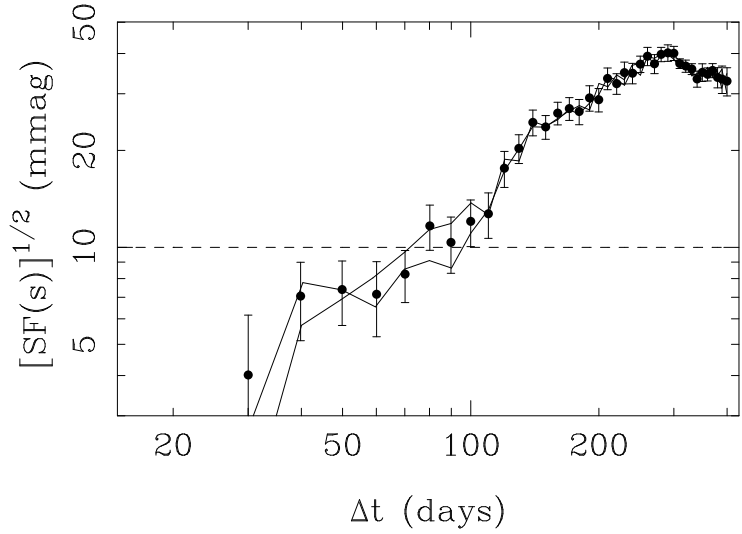}
\end{center}
\caption{First--order structure function of the underlaying intrinsic signal $s$ in the $r$ band. The 
$SF(s)$ from the $CLC$ in Fig. \ref{f3} (filled circles) is compared with two extreme structure functions associated 
with $CLC$s for the extreme values of the delay range (continuous lines). A discontinuous straight line 
represents the 10 mmag threshold.}
\label{f8}
\end{figure}

The $CLC$ is generated by intrinsic signal ($s$) and observational noise ($n$), so $y_{qso} = s + n$. This 
simple relationship between the combined fluxes, the underlaying intrinsic signal and the noise permits to 
estimate the first--order structure function of $s = y_{qso} - n$ \citep[e.g.][]{Gil01,Sim85}. A 
structure function analysis is a method of quantifying typical flux variabilities at different lags. The 
structure function $SF(s)$ at lag $\Delta t$ is given by
\begin{equation}
SF(s) = (1/2N) \sum_{i,j} [(y_{qso,j} - y_{qso,i})^2 - \sigma_i^2 - \sigma_j^2] ,
\end{equation}
where the sum only includes the ($i$,$j$) pairs verifying that $t_j - t_i \sim \Delta t$ (the number of 
such pairs is $N$). We take a normalization factor equal to 1/2, and thus, the asymptotic behaviour on long 
timescales is just the signal variance $\sigma_s^2$ instead of 2$\sigma_s^2$ \citep[e.g.][]{Col01}. Fig. \ref{f8}  
shows the structure function of \astrobj{SBS 0909+532} obtained from the $CLC$ and Eq. (1), using independent 10--day 
bins (filled circles). To test the robustness of the derived $SF(s)$, we also obtain extreme structure 
functions, i.e., using the extreme values of the delay range, $\Delta t_{BA}$ = $-$ 43 days and $\Delta t_{BA}$ 
= $-$ 55 days, to make the corresponding $CLC$s. These two extreme trends (continuous lines in Fig. \ref{f8}) are 
included in the error bars of the $SF(s)$, so the typical structure function (from $\Delta t_{BA}$ = $-$ 49 
days) is a reliable tool. The discontinuous line in Fig. \ref{f8} corresponds to a 10 mmag threshold, and at lags 
$\Delta t \leq$ 60 days, the typical fluctuations are below this 10 mmag level. This result explains why the 
flux ratio at the same time of observation basically coincides with the flux ratio corrected by the time delay, 
at least in the red arm of the optical spectrum \citep{Ull06}. 

\begin{figure}
\begin{center}
\includegraphics*[width=7cm,angle=-90]{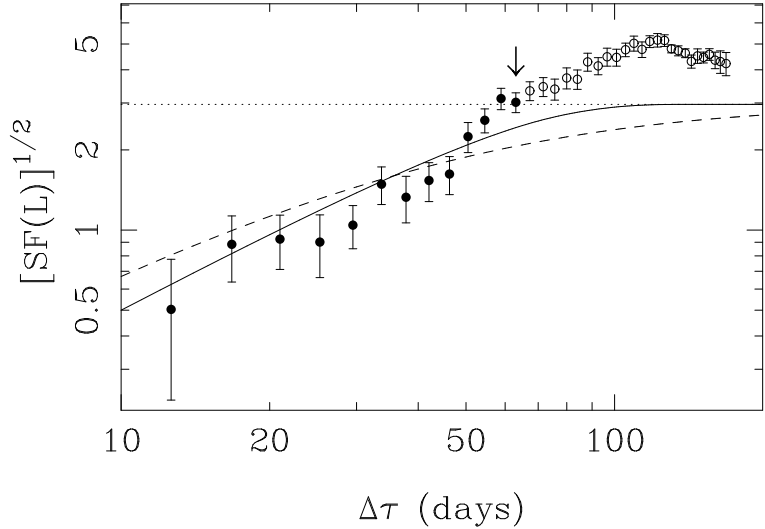}
\end{center}
\caption{Structure function of the intrinsic luminosity $L$. Taking into account the asymptotic behaviour 
(dotted line), the total monitoring period and the number of pairs in the bins, we distinguish between 
valid/reliable results (filled circles) and the rest of data points (open circles). The $SF(L)$ at $\Delta 
\tau >$ 63 days (rest--frame lags on the right of the arrow) is probably biased, and thus, the results at 
long rest--frame lags should be considered as unreliable data. While the solid line is the best fit from 
symmetric triangular flares in an accretion disc, the dashed line traces the best fit from standard nuclear 
starbursts (see main text).}
\label{f9}
\end{figure}

To discuss the origin of the intrinsic signal, we focus on the structure function of the intrinsic luminosity
$L$ \citep[e.g.][]{Cid00}. This $SF(L)$ can be directly compared to the predictions of different physical 
scenarios, e.g., nuclear starbursts and accretion disc instabilities \citep{Kaw98}. We remark that the observed 
flux at $\lambda$ is emitted at a shorter wavelength $\lambda/(1 + z_s)$, and we must consider the emission of 
near UV light ($\sim$ 2600 \AA), because the observations were made in the $r$ Sloan band ($\lambda \sim$  
6200 \AA). On the other hand, $s$ = $m_{qso} - m_b$ = $-$ 2.5 log($F_{qso}$/$F_b$), where $m$ and $F$ denote 
$r$--band magnitudes and monochromatic fluxes, respectively. We can also use the cosmological law $F_{qso} = 
\epsilon L_{qso}/[4\pi D_L^2 (1 + z_s)]$, with $L_{qso}$, $\epsilon$ and $D_L$ being respectively monochromatic 
luminosity of the quasar, extinction--magnification factor and luminosity distance. From some rearrangement, it 
is inferred the relationship $kL_{qso}$ = 10$^{- 0.4s}$, $k$ = $\epsilon/[4\pi D_L^2 (1 + z_s) F_b]$. We 
initially take a set of units so that $k$ = 1 and $L$ = $L_{qso}$ = 10$^{- 0.4 s}$, so the structure function 
$SF(L)$ at rest--frame lag $\Delta \tau$ = $\Delta t/(1 + z_s)$ can be estimated through the averaged sum
\begin{equation}
SF(L) = (1/2N) \sum_{i,j} [(10^{- 0.4 y_{qso,j}} - 10^{- 0.4 y_{qso,i}})^2 - \overline{\sigma}_i^2 - 
\overline{\sigma}_j^2] ,
\end{equation}
where $\overline{\sigma} = 0.921 \times 10^{- 0.4 y_{qso}} \sigma$ and the sum includes $N$ pairs verifying 
$\tau_j - \tau_i \sim \Delta \tau$. As the $[SF(L)]^{1/2}$ values from Eq. (2) are $\sim$ 10$^{-3}$, we decide 
to use more appropriate units: $k$ = 1/1000. In these new units, the structure function of the intrinsic 
luminosity is drawn in Fig. \ref{f9} (filled and open circles). In this figure, the dotted straight line represents 
the asymptotic behaviour $\sigma_L$, i.e., $[SF(L)]^{1/2} \to \sigma_L$ at long lags. From now on, we only 
consider the data points not exceeding the flat asymptotic behaviour (filled circles). There is an arrow to 
make mark on the last reliable point. This last valid point corresponds to $\Delta t \sim$ 150 days $\sim$ 
$P$/3. Moreover, at $\Delta t \leq$ 150 days, the number of pairs in each bin is $N \sim$ 100--200. Longer 
lags cannot be seriously considered in the analysis, since the associated information is probably biased and 
could not describe the typical behaviour of the signal. 

The reliable trend in Fig. \ref{f9} (filled points) should be related to the variability scenario and may unveil 
the origin of the intrinsic fluctuations. First, we use three simple analytical models of accretion disc 
instabilities. In these three Poissonian models, the luminosity is due to the superposition of a variable
component and a constant background (the nonflaring part of the accretion disc). The variable component is 
made by the superpositions of flares at random times, which are characterized by a typical timescale $T$.
Square (SQ), exponentially decaying (ED) and symmetric triangular (ST) flares are taken into account 
\citep[see Appendix B of][]{Cid00}. When the observed $SF(L)$ is fitted to the analytical intrinsic 
structure functions, the SQ and ED models do not work well and lead to $\chi^2 \sim$ 4--5. However, the ST
model leads to a more reasonable (but not sufficiently good) $\chi^2$ value of 2 (solid line in Fig. \ref{f9}). For
the best ST model, $T$ = 70 days and the lifetime is $\tau$ = 3$T$/2 $\sim$ 100 days. There are several 
physical timescales that might be associated with accretion disc instabilities. The main dynamical 
timescales are the free--fall time $\tau_{ff} \sim (R^3/GM)^{1/2}$ and the orbital time $\tau_{orb} \sim 
2\pi (R^3/GM)^{1/2} \sim 6 \tau_{ff}$. The thermal timescale is given by $\tau_{th} \sim \tau_{ff}/\gamma$, 
i.e., the time for vertical diffusion of heat. Apart from the black hole mass $M$ and the emission radius 
$R$, the thermal timescale also depends on the disc viscosity parameter $\gamma$ (this last parameter is 
usually named $\alpha$, but we rename it $\gamma$ to avoid confusion). The viscous time $\tau_{visc} \sim 
(R/h)^2 \tau_{th}$ ($h$ is the disc thickness) is also relevant. This is linked to the radial inflow of a 
gas element, and represents the longest timescale \citep[e.g.][and references therein]{Cze04,Kro99}. 
\citet{Col01} estimated the values of $\tau_{ff}$, $\tau_{orb}$ and $\tau_{th}$ for different values of $M$ 
at two rest--frame wavelengths: 1400 and 5000 \AA\ (see Fig. 5 in that work). To do the estimations, they 
assumed a standard geometrically thin and optically thick disc \citet{Sha73}. At $\sim$ 2600 \AA\ (see 
above), the lifetime of ST flares is consistent with the orbital time for $\sim$ 10$^9$ M$_\odot$ black 
holes and the thermal timescale for $\sim$ 10$^7$ M$_\odot$ black holes. 

Many previous studies focused on the logarithmic slope ($\beta$) of the growing part of the square root of
structure function, which (slope) is directly related to the physical mechanism responsible for the variability 
\citep[e.g.][]{Kaw98}. For example, for SQ flares of duration $T$, $[SF(L)]^{1/2} \propto (\Delta \tau)^{1/2}$ 
at lags $\Delta \tau \leq T$ ($SF(L)$ is flat at $\Delta \tau > T$). This leads to a constant slope $\beta$ = 
0.5. For ED and ST flares, $\beta$ reaches its maximum value at the shortest lags. If the ED flares are 
characterized by a semi--lifetime $T$, then $\beta \sim$ 0.5 at $\Delta \tau \leq T/2$. On the other hand, for 
ST flares with rise time $T$, $[SF(L)]^{1/2}$ is approximately proportional to $\Delta \tau$ at $\Delta \tau 
\leq T/2$. Thus, whereas $\beta$ should be less than or equal to 0.5 for the SQ and ED models, the ST model 
produces a steep slope $\beta \sim$ 1 at relatively short lags. What about the observed structure function?. 
The observed slope is $\beta \sim$ 1 using both the subset of data at $\Delta \tau \leq$ 35 days (the first 
six points in Fig. \ref{f9}) and the global data set until $\Delta \tau$ = 63 days. It is clear that the observed 
slope of about 1 favours the ST model of accretion disc instabilities. However, even ST flares with equal 
rise and decay times $T$ = 70 days (best solution) have not the ability to reproduce the observed slope at
lags exceeding 50 days (see Fig. \ref{f9}). In fact this is the reason to obtain a rough agreement ($\chi^2$ = 2) 
instead of an accurate fit leading to $\chi^2 \sim$ 1. The cellular--automaton model \citep{Min94} is another 
framework to describe accretion disc instabilities. In spite of its popularity, we cannot consider that 
physical mechanism as an alternative to explain the observations, since it generates a small slope: $\beta$ = 
0.4--0.5 \citep{Kaw98}.   

With respect to the nuclear starbursts, we use the standard model by \citet{Are97}. This simple model is 
characterized by a timescale $\tau_{sg}$, which is the time when the supernova remnant reaches the maximum of 
its radiative phase. The observed $SF(L)$ cannot be reproduced by the relationship in Eq. (17) of 
\citet{Are97}. Although the best timescale has an acceptable value of $\tau_{sg}$ = 60 days (considering that 
the source is a very bright quasar), the observations--model comparison leads to a poor fit with $\chi^2$ = 4 
(dashed line in Fig. \ref{f9}). Standard nuclear starbursts are able to induce a steep slope at lags shorter than 
$\tau_{sg}$. However, the flat regime is only reached at lags significantly longer than $\tau_{sg}$ 
\citep{Are97,Kaw98}. This mechanism is thus unable to reproduce a steep slope from short lags to the flattening
lags, and consequently it fails to trace the observed behaviour.   

\section{Conclusions and discussion}

The 2 m Liverpool Robotic Telescope \citep{Ste01,Ste04} at the Roque de los Muchachos Observatory (Canary Islands, 
Spain) is ideally suited to monitorize gravitationally lensed quasars (GLQs) and derive light curves of their 
components. In this paper we present the first GLQ monitoring campaign using the Liverpool Telescope. We have 
observed the double quasar \astrobj{SBS 0909+532} in 2005--2006, taking nightly frames in the $r$ Sloan filter. 
Our main results and conclusions are:
\begin{enumerate}
\item The $r$--band frames and PSF photometry lead to variable light curves of the two quasar components A and 
B. Both brightness records show a decline larger than 50 mmag followed by a 50--mmag event. These features are 
used to confirm the recently reported time delay between the components \citep{Ull06}. From different 
cross--correlation techniques and a large number of repetitions of the experiment (synthetic light curves based 
on the observed records), we infer a delay $\Delta t_{BA}$ = $-$ 49 $\pm$ 6 days (1$\sigma$ interval), which 
agrees with our previous results. The new delay determination is robust, since a wide range of possible delays 
is tested and the two light curves overlap over a long time interval, when the A record is shifted by the best 
solution of the time delay. 
\item To obtain the difference light curve of \astrobj{SBS 0909+532}, the magnitude-- and time--shifted light curve of 
image A is subtracted from the light curve of image B. There is no evidence in favor of the existence of 
extrinsic variability (e.g., fluctuations or gradients caused by microlensing in the lensing galaxy halo), since 
the difference curve is consistent with zero. Very recently, \citet{Par06} derived difference curves of five 
GLQs and found that two out of the five GLQs have significant extrinsic signal. \citet{Par06} also presented one 
difference curve that clearly agrees with zero (\astrobj{HE 2149--2745}) as well as two doubtful cases. Here we report on 
another GLQ with flat difference curve, which joins the family of systems without important extrinsic variations 
\citep[e.g.][]{Bur02,Gil01,Sch98}.  
\item Taking into account the absence or the extremely low (undetectable) level of extrinsic signal, we make the 
combined light curve and interprete it as due to the observational noise and intrinsic variations. The combined 
photometry consists of both A and B records, where the A light curve is shifted by the best solutions of the 
time delay and the magnitude offset. In order to study the growth of intrinsic variability with rest--frame lag, 
we explicitly obtain the structure function of the intrinsic luminosity. This is then fitted to predictions of 
simple models of accretion disc instabilities and nuclear starbursts. With respect to the disc--instability 
scenario, we concentrate on the phenomenological models by \citet{Cid00}, since the cellular--automaton model 
\citep{Min94} is related to a relatively small logarithmic slope of the structure function \citep{Kaw98}. This 
small slope ($\beta$ = 0.4--0.5) is not consistent with the structure function derived from observations ($\beta
\sim$ 1). We also consider the standard nuclear starbursts by \citet{Are97}. Symmetric triangular flares in an 
accretion disc lead to the best (but not accurate enough) fit, whereas standard SN explosions cannot produce 
a large value of $\beta$ from short to long lags, and thus, cannot account for the observed variability. 
\end{enumerate}

We obtain a rough agreement between the observations of \astrobj{SBS 0909+532} and the production of 100--day symmetric 
flares in the accretion disc of the distant source quasar. Considering typical supermassive black holes, the 
lifetime of the flares seems to agree with the dynamical and thermal times at the emission radius corresponding
to a standard disc \citep{Col01}. This suggest the existence of disc instabilities having the local (emission 
ring) dynamical--thermal timescale. Local magnetorotational instabilities \citep[e.g.][]{Bal91} would be possibly 
able to cause variability over this timescale. However, it is difficult to identify precisely the physical origin 
of the fluctuations, and we cannot rule out other processes. For example, the hottest temperature $kT$ at the 
innermost radius of a standard gas disc is clearly smaller than X--ray energies. Therefore, as \astrobj{SBS 0909+532} is a 
bright X--ray source \citep{Cha00,Pag04}, inverse Comptonization in a hot inner corona may explain its X--ray 
emission. Instabilities in the corona, which is presumably unstable \citep[e.g.][]{Sha76}, could lead to variable 
X--ray emission as well as variable X--ray irradiation of the gas disc. After X--ray reprocessing in the disc, 
variability at optical/UV wavelengths is expected \citep[e.g.][]{Cze04}. \citet{Man96} also proposed an unstable
advection--dominated disc that might generate X--ray flares and subsequent optical/UV events. In both physical
schemes, the shape and timescale of the optical/UV variations are determined by the shape and timescale of the 
X--ray fluctuations. Are these reasonable pictures?. X--ray symmetric flares were previously found by 
\citet{Neg94}, but on a stellar scale. \astrobj{Cygnus X--1} is an accreting black hole system ($M \sim$ 10 M$_\odot$) 
that produces symmetric flares lasting about 1 second \citep[see Fig. 1 of][]{Neg94}. The \astrobj{Cygnus X--1} variability 
was reproduced through theoretical calculations by \citet{Man96}, so it may be associated with 
advection--dominated disc instabilities. If the same mechanism is responsible for the X--ray variability of both 
stellar and supermassive black hole systems, then the stellar timescale must be rescaled by 10$^6$--10$^8$ 
(assuming typical supermassive black holes with $M \sim$ 10$^7$--10$^9$ M$_\odot$) to obtain $\tau \sim$ 10--1000 
days symmetric flares on a quasar scale. For $M \sim$ 10$^8$ M$_\odot$, these hypothetical X--ray events would be 
good tracers of our near UV fluctuations. Thus, surprisingly, we find that advection--dominated disc instabilities 
could be able to account for variations on both stellar and quasar scales. An unstable hot corona is also a 
plausible source of X--ray flares, but a discussion on the involved timescale and (a)symmetry is out of the scope
of this paper.

To date only one more GLQ with flat difference curve was analysed in some detail to determine the origin of its 
intrinsic variability. From light curves of \astrobj{QSO 0957+561} in two different optical bands ($gr$ filters), 
\citet{Kun97} 
inferred variability functions $[SF(m)]^{1/2}$ and their corresponding logarithmic slope $\beta_{gr} \sim$ 0.4. 
Here, $m = s + n$ are quasar magnitudes, i.e., the observational noise was not subtracted from the photometric
measurements. Using only data in the $g$ band, \citet{Kaw98} also determined a shallow slope of $[SF(m)]^{1/2}$:
$\beta_{g} \sim$ 0.3--0.4. \citet{Kaw98} presented a growing trend at lags $\Delta t \leq$ 500 days ($\Delta \tau 
\leq$ 200 days), but they did not report on the asymptotic behaviour ($\sqrt{2}\sigma_m$ for the Kawaguchi et al.'s 
normalization) or the reliability of the results at long lags, so we cannot properly discuss the involved 
variability timescale. On the other hand, the observed rise in the variability is consistent with a 
cellular--automaton disc--instability model \citep{Kaw98}. However, in spite of the small errors in the light curves
($\sigma \sim$ 10 mmag), the noise in the structure function ($\sqrt{2}\sigma \sim$ 15 mmag) does have a significant 
effect on the measured variations at the shortest lags and the measured sope. Thus, the slope of $[SF(s)]^{1/2}$ 
should be steeper than the slope of $[SF(m)]^{1/2}$. If we roughly construct a noise--less function $[SF(s)]^{1/2}$
in the $g$ band, the new slope is $\sim$ 0.7. Moreover, this new (and probably true) slope agrees with the slope in 
the $R$ band \citep{Gil01}. Therefore, there are clear evidences that both \astrobj{QSO 0957+561} and 
\astrobj{SBS 0909+532} are $z$ = 1.4 bright quasars with $\beta \sim$ 0.7--1. 

We may also try to establish a relation between these GLQ results and variability studies of non--lensed quasars. 
While some works focused on ensemble structure functions of large samples (hundreds, thousands or even more) of 
quasars \citep[e.g.][]{deV05,Haw02,Van04}, other works cocentrated on well--sampled individual quasars 
\citep[e.g.][]{Cid00,Col01}. The first ones did not include details on individual objects. Moreover, additional 
problems usually complicate a direct comparison with current GLQ data. For example, \citet{deV05} stated that their 
study is insensitive (due to the measurement noise) to lags shorter than $\tau \sim$ 1 year. Thus, we cannot put the 
behaviour of GLQs in perspective. \citet{Haw02} mostly used relatively faint objects showing coherent fluctuations of 
about 1 mag over a few years (see Fig. 4 in that work). However, both GLQs are relatively bright sources displaying 
coherent flux variations of about 0.1 mag over a few or several months. On timescales of a few years, the GLQs have a 
maximum scatter (difference between maximum and minimum flux) below 0.25 mag. Moreover, instead of a noise--less 
structure function having a good time resolution at lags $\Delta t \leq$ 1--2 years, \citet{Haw02} studied a long term 
ensemble $[SF(m)]^{1/2}$ with time resolution of about one year. The most relevant ensemble variability was reported by 
\citet{Van04}, who showed an $[SF(s)]^{1/2}$ vs. $\Delta \tau$ relationship having reasonable time coverage and 
resolution. This ensemble variability for quasars with very different luminosities (absolute magnitudes) and redshifts 
led to a slope $\beta \sim$ 0.3 at $\Delta \tau \leq$ 500 days. It is evident that the GLQ slope $\beta >$ 0.5 
disagrees with the ensemble slope ($\beta <$ 0.5), which could indicate the existence of different populations of 
intrinsically variable quasars (perhaps depending on luminosity and redshift) or even the existence of microlensing 
effects on a large amount of quasars \citep[e.g.][]{Haw02}. With respect to the well--sampled individual sources, 
\citet{Cid00} analysed the light curves of 42 $z <$ 0.4 quasars that were monitored by the Wise Observatory group 
\citep{Giv99}. Assuming the presence of square flares ($\beta$ = 0.5), they derived lifetimes ranging from 124 days to 
a few years, so the shortest ones are close to the lifetime of the \astrobj{SBS 0909+532} flares. \citet{Cid00} also 
remarked 
the insensitivity of their fits to the flare shape (with the exception of exponentially decaying flares, since these 
asymmetric flares led to poor fits). These results (relatively long lifetimes and poor sensitivity to the flare shape) 
could be due, at least in part, to the procedure for fitting observations, which involved two free parameters, i.e., 
the flare lifetime and the variance. We, however, compute directly the variance from the light curve and then fit the 
lifetime. Our variance represents the typical variability in a period of about 200 days (rest--frame time), and 
taking into account the absence of important gradients over 1.5 years \citep[from a comparison with previous Calar 
Alto light curves][]{Ull06}, this seems a good tracer of the variance over longer periods. \citet{Col01} studied 13 
local active galactic nuclei with very good time coverage and resolution. The slopes of their optical/UV structure 
functions varied between $\beta \sim$ 0.3 and $\beta \sim$ 0.8 (similar to the GLQ slope). The flare lifetimes (using 
symmetric triangular flares and certain lag intervals) are $\tau \sim$ 5--94 days, with the longest ones very close to 
the lifetime of the \astrobj{SBS 0909+532} events. In order to check the results based on a monitoring period of 200--300 
days, 
\citet{Col01} presented the variability structure for $\sim$ 7 years of monitoring of \astrobj{NGC 5548}. Using the short 
monitoring period, they found $\tau \sim$ 40--60 days. However, from the additional analysis of the longer monitoring 
period, they inferred richer results. The asymptotic behaviour is reached at lags of $\sim$ 200 days, and the 50--day 
timescale seems to represent the lifetime of the shortest flares.   
    
To accurately reproduce the observed variability of \astrobj{SBS 0909+532}, new (more complex) models seem to be 
required. Moreover, a longer monitoring period would lead to a very precise description of the structure at $\Delta 
\tau \leq$ 60 days and the appearance of reliable features at lags exceeding $\Delta \tau \sim$ 60--70 days, which 
would significantly improve the observational constraints. We also note that the current optical data of 
\astrobj{SBS 0909+532}
are insufficient to carry out more ambitious analyses, e.g., a test for non--linearity. \citet{Utt05} showed that a 
linear variability--flux relation in the light curve of an object is an indicator of non--linearity. This hypothetical 
linear relation would suggest that the variability process is multiplicative, or in other words, it is the result of 
several random subprocesses which multiply together. A process produced by multiplication of many independent processes 
has a lognormal distribution, so the hypothetical light curve would have a lognormal probability density function (PDF)
\citep[for details, e.g.][]{Utt05}. However, the observed fluctuations in the flux of \astrobj{SBS 0909+532} ($\leq$ 10\%) 
and the number of available data points (clearly insufficient to make highly populated bins in flux) do not permit to 
reliably obtain the variability--flux relation or the PDF of \astrobj{SBS 0909+532}. Finally, we remark the unique 
advantages of 
using GLQs as a tool to study the origin of the intrinsic signal of quasars, because there is no way to disentangle 
intrinsic from extrinsic signal in a non--lensed quasar. New long term monitoring programmes of GLQs with flat 
difference curve will permit to fairly discuss the structure of the intrinsic variability on short and long timescales, 
and to compare between this behaviour and the structure of the fluctuations in both GLQs with non--flat difference 
curve and non--lensed quasars.

\section*{Acknowledgments}

We thank an anonymous referee for several comments which significantly improved the discussion concerning the 
intrinsic variability. We are also indebted to C. Moss for guidance in the preparation of the robotic monitoring 
programmes. To check the monitoring campaign in Spain, we use observations at Mt. Maidanak (Uzbekistan). We 
acknowledge the observers and heads of the Maidanak collaboration for their maintenance and availability of the 
gravitational lenses database. This research has been supported by the Spanish Department of Education and Science 
grant AYA2004-08243-C03-02, University of Cantabria funds, grant for young scientists of the President of the 
Russian Federation (number MK-2637.2006.2), Deutscher Akademischer Austausch Dienst (DAAD) grant number A/05/56557, 
grant of Russian Foundation for Basic Research (RFBR) 06-02-16857 and the Science and Technology Center of Ukraine 
(STCU) grant U127k. We also acknowledge support by the European Community's Sixth Framework Marie Curie Research 
Training Network Programme, Contract No.MRTN-CT-2004-505183 "ANGLES". Funding for the Sloan Digital Sky Survey 
(SDSS) has been provided by the Alfred P. Sloan Foundation, the Participating Institutions, the NASA, the NSF, the 
U.S. Department of Energy, the Japanese Monbukagakusho, and the Max Planck Society. The SDSS is managed by the 
Astrophysical Research Consortium (ARC) for the Participating Institutions. The Participating Institutions are The 
University of Chicago, Fermilab, the Institute for Advanced Study, the Japan Participation Group, The Johns Hopkins 
University, Los Alamos National Laboratory, the Max-Planck-Institute for Astronomy (MPIA), the Max-Planck-Institute 
for Astrophysics (MPA), New Mexico State University, University of Pittsburgh, Princeton University, the United 
States Naval Observatory, and the University of Washington.

\clearpage

\begin{center}
\begin{longtable}{ccccc} 
\caption{LRT fluxes of \astrobj{SBS 0909+532}A,B in the $r$ Sloan filter} \label{t2}\\
\hline \multicolumn{1}{c}{JD$-$2450000} & 
\multicolumn{1}{c}{$m_A$ (mag)} & 
\multicolumn{1}{c}{$\sigma(m_A)$ (mag)} &
\multicolumn{1}{c}{$m_B$ (mag)} & 
\multicolumn{1}{c}{$\sigma(m_B)$ (mag)} \\
\hline 
\endfirsthead 

\multicolumn{5}{c}%
{\tablename\ \thetable{}: LRT fluxes of \astrobj{SBS 0909+532}A,B in the $r$ Sloan filter
(continuation)} \\ 
\hline \multicolumn{1}{c}{JD$-$2450000} & 
\multicolumn{1}{c}{$m_A$ (mag)} & 
\multicolumn{1}{c}{$\sigma(m_A)$ (mag)} &
\multicolumn{1}{c}{$m_B$ (mag)} & 
\multicolumn{1}{c}{$\sigma(m_B)$ (mag)} \\
\hline 
\endhead 

   3446.388   &    16.337  &    0.014 &  16.985  &    0.018   \\
   3453.428   &    16.336  &    0.014 &  16.984  &    0.018   \\
   3455.432   &    16.349  &    0.014 &  16.977  &    0.018   \\
   3458.426   &    16.342  &    0.014 &  16.981  &    0.018   \\
   3487.371   &    16.333  &    0.014 &  16.964  &    0.018   \\
   3489.371   &    16.321  &    0.014 &  16.970  &    0.018   \\
   3503.383   &    16.312  &    0.014 &  16.978  &    0.018   \\
   3509.389   &    16.325  &    0.014 &  16.965  &    0.018   \\
   3510.384   &    16.303  &    0.014 &  16.989  &    0.018   \\
   3522.463   &    16.318  &    0.014 &  16.961  &    0.018   \\
   3525.425   &    16.317  &    0.014 &  16.969  &    0.018   \\
   3646.719   &    16.380  &    0.014 &  17.060  &    0.018   \\
   3649.711   &    16.383  &    0.010 &  17.067  &    0.013   \\
   3656.713   &    16.395  &    0.010 &  17.047  &    0.013   \\
   3662.709   &    16.386  &    0.010 &  17.051  &    0.013   \\
   3667.711   &    16.382  &    0.014 &  17.074  &    0.018   \\
   3676.662   &    16.395  &    0.010 &  17.025  &    0.013   \\
   3678.723   &    16.390  &    0.014 &  17.032  &    0.018   \\
   3680.637   &    16.392  &    0.014 &  17.026  &    0.018   \\
   3681.645   &    16.391  &    0.014 &  17.044  &    0.018   \\
   3682.660   &    16.383  &    0.014 &  17.047  &    0.018   \\
   3684.670   &    16.392  &    0.010 &  17.052  &    0.013   \\
   3693.615   &    16.390  &    0.010 &  17.042  &    0.013   \\
   3699.588   &    16.396  &    0.010 &  17.051  &    0.013   \\
   3700.680   &    16.398  &    0.014 &  17.006  &    0.018   \\
   3707.590   &    16.399  &    0.014 &  17.007  &    0.018   \\
   3708.576   &    16.390  &    0.010 &  17.048  &    0.013   \\
   3709.699   &    16.394  &    0.010 &  17.030  &    0.013   \\
   3711.699   &    16.389  &    0.014 &  17.050  &    0.018   \\
   3712.621   &    16.389  &    0.014 &  17.044  &    0.018   \\
   3714.590   &    16.397  &    0.010 &  17.048  &    0.013   \\
   3718.555   &    16.396  &    0.010 &  17.037  &    0.013   \\
   3719.615   &    16.391  &    0.010 &  17.039  &    0.013   \\
   3728.629   &    16.402  &    0.014 &  17.037  &    0.018   \\
   3729.504   &    16.386  &    0.014 &  17.056  &    0.018   \\
   3731.544   &    16.394  &    0.008 &  17.031  &    0.010   \\
   3766.609   &    16.405  &    0.014 &  17.030  &    0.018   \\
   3783.531   &    16.403  &    0.014 &  17.014  &    0.018   \\
   3784.539   &    16.399  &    0.014 &  17.018  &    0.018   \\
   3787.457   &    16.401  &    0.014 &  17.025  &    0.018   \\
   3789.449   &    16.396  &    0.014 &  17.027  &    0.018   \\
   3790.422   &    16.400  &    0.014 &  17.024  &    0.018   \\
   3793.410   &    16.392  &    0.014 &  17.015  &    0.018   \\
   3798.445   &    16.406  &    0.014 &  16.987  &    0.018   \\
   3802.406   &    16.390  &    0.014 &  17.009  &    0.018   \\
   3803.395   &    16.391  &    0.014 &  17.002  &    0.018   \\
   3804.418   &    16.391  &    0.014 &  17.018  &    0.018   \\
   3805.391   &    16.402  &    0.014 &  17.004  &    0.018   \\
   3806.383   &    16.387  &    0.014 &  17.009  &    0.018   \\
   3813.395   &    16.382  &    0.014 &  17.019  &    0.018   \\
   3814.426   &    16.382  &    0.014 &  17.014  &    0.018   \\
   3815.383   &    16.381  &    0.014 &  17.009  &    0.018   \\
   3817.438   &    16.388  &    0.014 &  17.034  &    0.018   \\
   3818.426   &    16.373  &    0.014 &  17.032  &    0.018   \\
   3820.379   &    16.384  &    0.014 &  17.024  &    0.018   \\
   3821.492   &    16.376  &    0.014 &  17.039  &    0.018   \\
   3822.492   &    16.390  &    0.014 &  17.028  &    0.018   \\
   3826.492   &    16.379  &    0.014 &  17.012  &    0.018   \\
   3827.520   &    16.380  &    0.014 &  17.038  &    0.018   \\
   3828.492   &    16.364  &    0.014 &  17.032  &    0.018   \\
   3830.473   &    16.384  &    0.014 &  17.018  &    0.018   \\
   3833.500   &    16.368  &    0.014 &  17.036  &    0.018   \\
   3841.465   &    16.354  &    0.014 &  17.063  &    0.018   \\
   3843.430   &    16.368  &    0.014 &  17.058  &    0.018   \\
   3845.430   &    16.367  &    0.014 &  17.030  &    0.018   \\
   3854.402   &    16.382  &    0.014 &  17.028  &    0.018   \\
   3855.488   &    16.370  &    0.014 &  17.034  &    0.018   \\
   3856.402   &    16.384  &    0.014 &  17.029  &    0.018   \\
   3860.391   &    16.397  &    0.014 &  17.026  &    0.018   \\
   3862.395   &    16.386  &    0.014 &  17.035  &    0.018   \\
   3864.453   &    16.378  &    0.014 &  17.037  &    0.018   \\
   3867.438   &    16.380  &    0.014 &  17.041  &    0.018   \\
   3870.434   &    16.379  &    0.014 &  17.042  &    0.018   \\
   3875.418   &    16.372  &    0.014 &  17.057  &    0.018   \\
   3876.477   &    16.362  &    0.014 &  17.036  &    0.018   \\
   3879.430   &    16.378  &    0.014 &  17.030  &    0.018   \\
   3880.414   &    16.361  &    0.014 &  17.039  &    0.018   \\
   3888.418   &    16.369  &    0.014 &  17.025  &    0.018   \\
\hline
\end{longtable}
\end{center}

\end{document}